\documentclass[a4paper,11pt]{article}
\usepackage{jinstpub} 
\usepackage{threeparttable}
\usepackage{silence}

\title{LET measurements and simulation modelling of the charged particle field for the Clatterbridge ocular proton therapy beamline}

\author[a,b,c,1]{Jacinta S.L. Yap,\note{Corresponding author.}}
\author[d,e,f]{Navrit J.S. Bal,}
\author[g]{Mark D. Brooke,}
\author[h]{Cristina Oancea,}
\author[i,h]{Carlos Granja,}
\author[j,k]{Andrzej Kacperek,}
\author[k]{Simon Jolly,}
\author[g]{Frank Van den Heuvel,}
\author[l]{Jason L. Parsons}
\author[b,c]{and Carsten P. Welsch}
\affiliation[a]{School of Physics, University of Melbourne, Melbourne, Australia}
\affiliation[b]{University of Liverpool, Liverpool, United Kingdom}
\affiliation[c]{Cockcroft Institute, Warrington, United Kingdom}
\affiliation[d]{Nikhef, Amsterdam, Netherlands}
\affiliation[e]{Department of Clinical Medicine, Aarhus University, Aarhus, Denmark}
\affiliation[f]{Danish Centre for Particle Therapy, Aarhus University Hospital, Aarhus, Denmark}
\affiliation[g]{Department of Oncology, University of Oxford, Oxford, United Kingdom}
\affiliation[h]{ADVACAM, Prague, Czech Republic}
\affiliation[i]{VSB -- Technical University of Ostrava, Ostrava, Czech Republic}
\affiliation[j]{The Clatterbridge Cancer Centre NHS Foundation Trust, Wirral, United Kingdom}
\affiliation[k]{University College London, University of London, London, United Kingdom}
\affiliation[l]{Department of Cancer and Genomic Sciences, University of Birmingham, Birmingham, United Kingdom}
\emailAdd{jacinta.yap@unimelb.edu.au}

\abstract{Proton therapy can achieve a highly targeted treatment by utilising the advantageous dosimetric characteristics of the Bragg Peak. Protons traversing through a material will deposit their maximum energy at the Bragg Peak through ionisation and other interactions, transferring minimal excess dose to surrounding tissue and organs. This rate of energy loss is also quantified by the linear energy transfer (LET), which is indicative of radiation quality and radiobiological effects. However it is a challenging physical quantity to measure, as characterisation of radiation fields and the impact of LET on treatment requires advanced tools and technology. The MiniPIX-Timepix is a miniaturised, hybrid semiconductor pixel detector capable of high resolution spectrometric tracking, enabling wide-range detection of the deposited energy, position and direction of single particles. Experimental measurements were performed at a clinical facility, the Clatterbridge Cancer Centre which houses a 60~MeV ocular proton therapy beamline. A realistic end-to-end model of the facility was developed in the Monte Carlo code TOPAS (TOol for PArticle Simulation) and was used to simulate the experimental conditions. The detector was held at 45$^{\circ}$ and 60$^{\circ}$ perpendicular to the beam, and placed downstream of various thickness Polymethyl methacrylate (PMMA) blocks to acquire data along the dose deposition depth. Empirical cluster data providing track length and the energy deposition distributions were used to obtain the LET spectra. The determined values for the LET in silicon and dose averaged LET across the BP show general agreement with simulated results, supporting the applicability of the TOPAS CCC model. This work explores the capability of the MiniPIX detector to measure physical quantities to resolve the LET, and discusses experimental considerations and further possibilities.}


\begin{document}
\WarningFilter{hyperref}{Ignoring empty anchor}
\maketitle
\flushbottom

\section{Introduction}
The Clatterbridge Cancer Centre (CCC) in the UK is the world’s first hospital based proton beam therapy (PBT) facility and has provided a successful treatment service for ocular tumours for more than 30 years \cite{Kacperek2009}. The clinic treats with a 60~MeV proton beam at isocentre, shaped precisely for each patient treatment with a passive double scattered delivery system. High rates of local tumour control, ocular retention and preservation of visual acuity have been achieved as the sites are well suited for this method of treatment \cite{Damato2005}. PBT has become an established treatment for ocular cases (uveal melanomas) among other modalities \cite{Mishra2016}.

Modern PBT clinics (typically high energy, multi-room and with gantries) are vendor built however there are many dedicated ocular proton beamlines worldwide. The CCC is a pioneering and unique facility which has supported a broad array of research and experimental work over the years \cite{Chaudhary2016,Vitti2019,Yap2021b,Kelleter2020,Taylor2016a}. To fully exploit the beamline, several simulation codes were used to characterise and accurately model the facility \cite{Yap2020}. The CCC treatment line was originally developed using the Monte Carlo (MC) simulation toolkit, Geant4 \cite{Agostinelli2003} and a successive version was recreated using TOol for PArticle Simulation (TOPAS) \cite{Perl2012,Faddegon2020}, a widely used medical physics code developed specifically for research and clinical application in PBT. Redeveloping the beamline in TOPAS enabled further capabilities: these developments and the use of the model is described in this work.  

MC simulations provide the most accurate predictions of interactions and calculations of dose relevant to treatment. However, they are computationally demanding and often require resources and processing times which can be impractical for routine clinical use. Analytical and even efficient MC algorithms are commonly used in PBT commercial treatment planning systems but can be inaccurate or insufficient for certain cases, such as the implementation of biologically weighted treatment quantities to represent the changes and uncertainties of biological effects \cite{McMahon2018}. A quantitative representation of radiation quality to compare the potential biological impact of two different ionising radiation modalities is known as the `\textit{Relative Biological Effect}' (RBE). In typical clinical practice, a constant RBE value of 1.1 is used for protons, meaning the physical dose is considered 10\% more effective compared with conventional X-ray radiation therapy. This is primarily due to the `Bragg Peak' (BP), which results in greater localised energy transfer, increased cell kill efficiency, and therapeutic benefit. There is ongoing debate over the consideration of approximating RBE as a single value given its dependence on numerous parameters and variability along the particle range, particularly across the spread out BP, at the BP and the distal fall-off \cite{Paganetti2014,Paganetti2019,Sorensen2021} -- this is however not further discussed here as it is beyond the scope of this work.

Instead, in the context of dose and implications for treatment, we examine a parameter which relates a physical quantity to radiobiological effects: the linear energy transfer. LET correlates cellular damage as a result of energy losses (d\textit{E}) along a track segment (d$\ell$), typically reported in units of keV/$\upmu$m. This value can change across the total path length, according to the particle type. LET is determined from electronic interactions with matter \cite{Seltzer2011} where protons and heavier charged particles are more damaging due to the increased cross section of electronic interactions and energy losses at the end of range. This increased density of ionisation results in higher LET values compared to photon induced interactions, where secondary electrons are generated with lower LET contributions \cite{VanDenHeuvel2014}. LET is similar to restricted stopping power which generally does not include contributions from nuclear interactions or secondary electrons which leave the track, transferring their energy elsewhere. There are several definitions of LET, as there are many factors \cite{Smith2021,Kalholm2021} which impact how to adequately calculate and score it either analytically, or with Monte Carlo methods \cite{Wilkens2004,Wilkens2003,Romano2014,Cortes-Giraldo2015,Guan2015}; in its canonical form, 
\begin{equation}
    \mathrm{LET} = \frac{\mathrm{d}\mathrm{E}}{\mathrm{d}\ell}.
\label{eqn1}
\end{equation}
As primary and secondary particles will have a range of resulting energies, their dose contributions can be weighted to provide a single, generalised LET metric: the dose-averaged LET (LET\textsubscript{d}) \cite{Bertolet2020}. LET\textsubscript{d} is also considered as an indicator of biological effects as caused by particle interactions, and has been shown as a predictor of the RBE for protons \cite{Chaudhary2014,Grassberger2011,Guan2018}. This is also meaningful for the radiobiological work performed on the CCC beamline which included cell studies investigating induced cellular responses, damage and repair between proton and photon beam radiation \cite{Fabbrizi2024,Vitti2019,Carter2019,Vitti2020,Nickson2021}. However, it is still not yet completely understood what would best represent the variation in beam quality for clinical treatment optimisation, or if LET or related quantities are sufficiently reliable \cite{Grun2019,Guan2024}; in this paper we contribute our LET\textsubscript{d} work for its potential relevance and application. 

Furthermore, in practice it is difficult to measure LET due to performance requirements which exceed the typical capabilities of commonly used methods of detection. Primarily, the position, charge and distribution of tracks deposited by individual particles must be able to be rapidly recorded and resolved with very high spatial resolution. To provide these measurements under these conditions is challenging and has not been realised for existing commercially available systems. Several studies with novel silicon detectors (Medipix, Timepix, VELO etc.) and micro-dosimeters have shown promise for quality assurance and other applications for radiotherapy \cite{Rosenfeld2020,Yap2021b,Bal2021,Schnuerer2018a}. Proceeding from developments made by the Medipix Collaboration (CERN), the Timepix \cite{Llopart2007} chip has been previously explored for applications in ion beam therapy, demonstrating the capacity to precisely measure LET relevant quantities such as the track length and deposited energy along the particle track, in different radiation environments \cite{Harrison2024,Oancea2023,Novak2023,Granja2021,Hoang2012,Opalka2012,Opalka2013,Jakubek2011,Jakubek2013,Stoffle2018}. In this paper we present details of an end-to-end, realistic simulation model of the Clatterbridge ocular beamline developed in TOPAS and its application for LET modelling and measurements of the 60~MeV protons. We use a MiniPIX detector to experimentally measure relevant quantities to resolve and obtain the LET spectra in silicon at different positions around the BP. 

\section{Methods}
\subsection{Clatterbridge Beamline TOPAS Model}
A complete end-to-end model of the Clatterbridge proton therapy beamline was developed in TOPAS as a simulation platform for experimental and biological work done at the facility. Following the implementation of the precise geometry of all components, its performance was benchmarked and matched with quality assurance data \cite{Yap_Thesis}. The model was built using TOPAS v3.2 but updated for compatibility with newer versions of TOPAS; it remains publicly accessible for all users as documented on \cite{UCLTOPAS}.

\subsubsection{Component Geometry}
\begin{figure}[!htb]
    \centering  
    \includegraphics*[width=0.74\columnwidth]{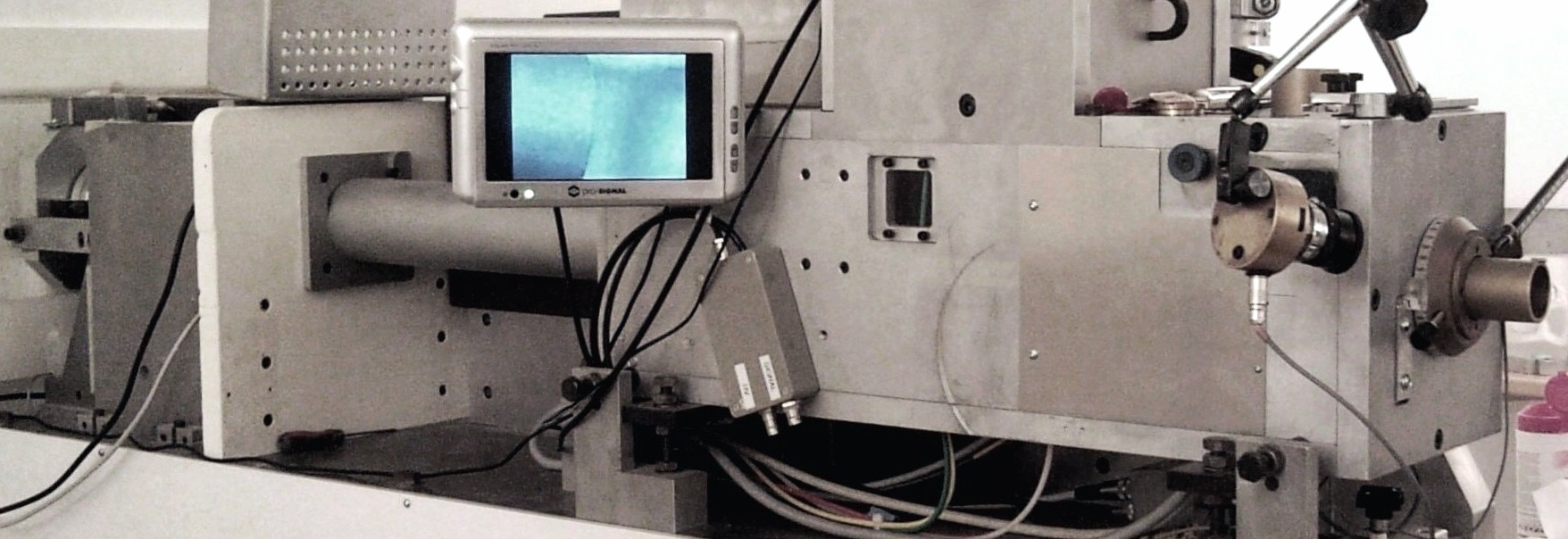}
    \includegraphics*[width=0.77\columnwidth]{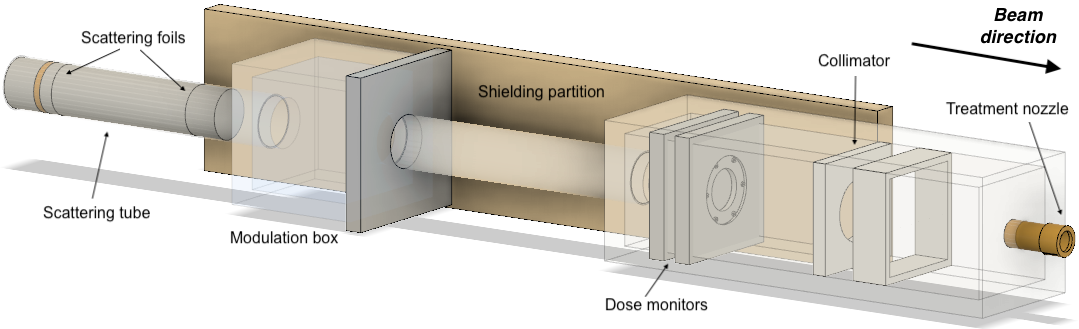}
    \caption{Image of the CCC treatment line from the wall of the treatment room to the patient nozzle (top). Delivery components downstream of the wall are enclosed but exposed to air. 3D rendering of the treatment line in CAD (bottom), components and direction of beam labelled.}
    \label{fig1}
\end{figure}

As CCC produces a passively scattered beam, the dimensions of all components needed to be verified to build an accurately representative model of the treatment line. Each component was physically remeasured ($\pm$0.5~mm precision \cite{UCLmaps}) and implemented as design objects built using CAD software (figure \ref{fig1}). 

Schematics were obtained and used to describe components which were not able to be measured in practice, specifically the dose monitors (developed in-house). Complex geometries can be constructed in TOPAS by using the TsCAD (TOPAS CAD) import feature which enables the conversion of any CAD (Computer Aided Design) generated objects into TOPAS geometry by Stereolithography binary (STL) or Polygon ASCII (PLY) format. The majority of the components were imported as STL files using this feature to preserve precise details of the treatment line system. However, several remaining components needed to be user defined in TOPAS individually, as they were unable to be correctly imported and presented issues or difficulties due to their geometry.

\begin{figure}[!htb]
\centering
\includegraphics[width=\columnwidth]{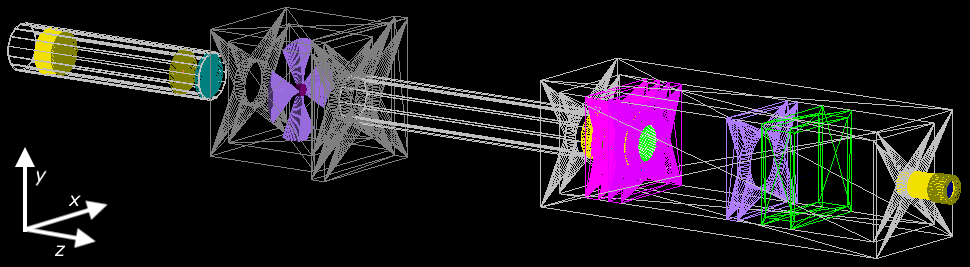}
\includegraphics[width=\columnwidth]{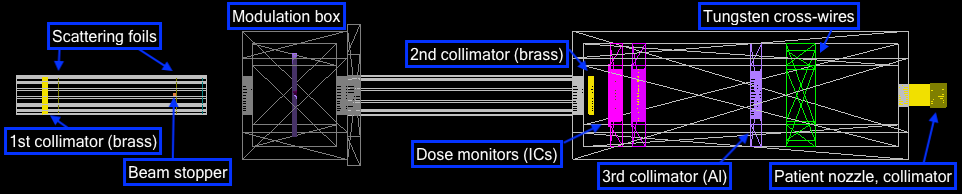}
\caption[]{CCC treatment beamline model visualised in TOPAS (top), wireframe view with global axes (beam direction, +$z$). Bird's-eye view with major components labelled (bottom).} 
\label{fig2}
\end{figure}

As STL only defines the geometry, every component needed to be grouped by material and imported as individual STL files. Each component group also has its own respective axis within the world coordinate system, retained during export. As TOPAS does not automatically determine the origin point, it may randomly place the imported component either at the centre or the edge of the parent volume. To mitigate this, some elements required additional rotation, translation and adjustment to reconcile the different coordinate systems. Clearances were also built between each adjacent surface to provide a 50~$\upmu$m tolerance between each component as TOPAS does not allow overlapping structures. This was chosen as a reasonable separation distance between elements to avoid simulation problems with stuck tracks and errors from geometry issues. The realistic model of the CCC treatment line developed in TOPAS incorporating the exact geometrical measurements of each component is shown in figure \ref{fig2}.

The main components of the treatment line comprise the first vacuum tube containing the double scattering tungsten foils and beam stopper, aluminium modulation box for range shifters or absorbers, dosimetry box with several collimators and custom-built diagnostics, and the patient nozzle. Following the beam stopper the proton beam exits vacuum through a kapton window and enters air. The delivery system components combine to minimise the beam penumbra and generate an output beam distribution suitable for patient treatments.

The CCC beam is produced upstream by a cyclotron. The beam source was defined at the exit of the cyclotron by its geometrical (size and angular divergence) properties and the Twiss parameters. This was determined from previous work \cite{Yap2020} given the historical changes and an absence of diagnostics upstream of the delivery system to provide a realistic description of the beam at present day conditions. 

\subsubsection{Simulation Conditions}
Simulations were run with the CCC model using TOPAS 3.2.p1 with standard (benchmarked) beam conditions where the pristine BP occurs at 30.5~mm. For a reference case, a beam of $10^6$ protons was generated and scored across a $35\times 35 \times 35~$mm$^{3}$ water phantom to obtain the proton depth dose profile and corresponding LET\textsubscript{d} values. A standard volume scorer was used for the dose profile and the prebuilt TOPAS `\textit{ProtonLET}' scorer to obtain LET\textsubscript{d} values. These are scored and calculated according to verified methods discussed in \cite{Granville2015,Cortes-Giraldo2015}.

\begin{figure}[!htb]
    \centering  
    \includegraphics*[width=0.65\columnwidth]{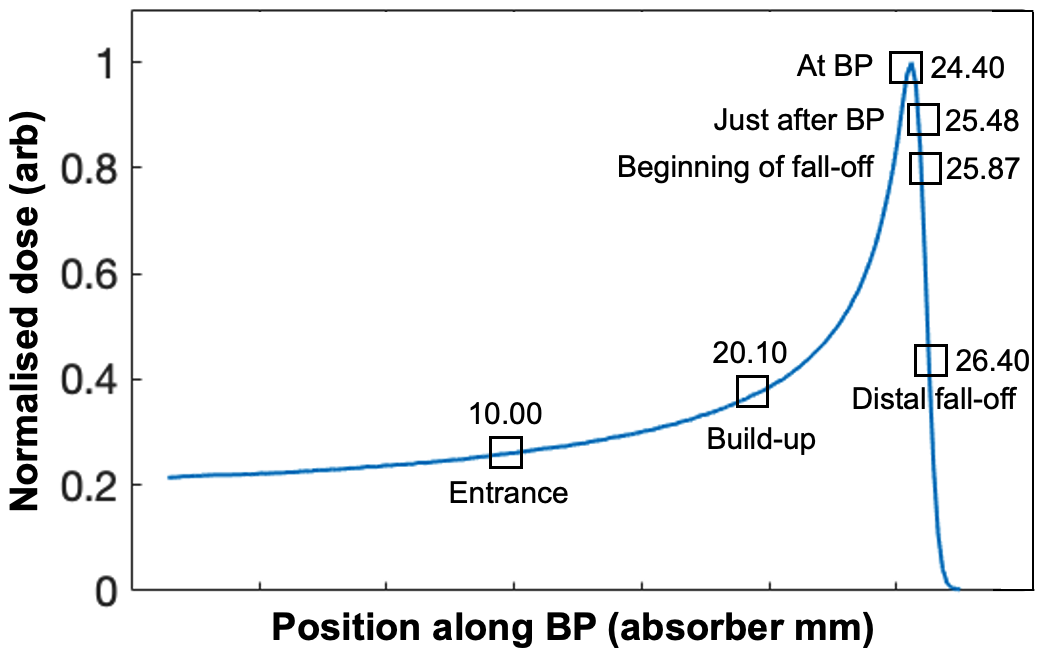}
    \caption{Graph showing the simulated CCC dose profile with points of interest denoted and corresponding thickness (mm) of PMMA absorbers required.}
    \label{fig3}
\end{figure}

A range of varying thickness PMMA absorbers were implemented in the simulation to determine the expected proton BP depths necessary to shift the BP relative to the sensor location. This identified the approximate range of PMMA block thicknesses needed for the experiment, to be able to probe at meaningful points around the pristine BP (figure \ref{fig3}): entrance region, build-up region, at BP, just after BP, start of fall-off, and distal fall-off. The geometry of the silicon sensor (with aluminium sliding cover) were also built in the model to simulate the exact experimental conditions for direct comparison with measured results. 

As only a small number of particles ultimately reach the sensor due to low transmission, a phase space scorer was implemented after the nozzle and used as the particle source. Subsequently, an output file of $\sim$1 million histories, showing a Gaussian beam with mean energy of 60.04~MeV and energy spread of 0.48~MeV was used for the experimental simulations. 

The LET\textsubscript{d} was scored on a 28 $\times$ 28 grid of 784 bins to represent the MiniPIX silicon sensor, each with a depth in the $z$-dimension of 300~$\upmu$m. This provided suitable statistics for comparison as cluster sizes would mostly trigger up to $\sim$40 pixels in the sensor, corresponding to an approximate resolution of a single cluster per bin. This was determined to be an appropriate trade off between cluster resolution and computational demand.

\subsection{MiniPIX-Timepix Detector}
The MiniPIX detector is a compact, hybrid semiconductor pixel detector with a Timepix ASIC with a 300~$\upmu$m silicon sensor and 14~$\times$~14~mm$^2$ active area. This encompasses a 256~$\times$~256 pixel array with 65,536 independent channels and readout via USB at up to 45 frames per second (FPS). The Timepix technology provides per-pixel signal processing in wide range for precise counting, energy or timing at the pixel level: quantum imaging sensitivity and spectrometric tracking provide fluxes and dose rates to resolve the beam profile, time, spatial dose mapping and LET (0.01~--~>100~keV/$\upmu$m) \cite{Granja2018,Granja2018b}. Various detection and analysis techniques have also been further developed to better identify clusters and determine particle data, these are not discussed in this paper but more is detailed in \cite{Martisikova2011,Granja2016,Granja2018a,Stasica2023,Nabha2022}. 

The detector can image single particle tracks in high resolution with 100\% collection efficiency from the size of the pixel (55~$\upmu$m) up to the sensor thickness (300~$\upmu$m). The absorption depth can be changed depending on the angle the sensor is positioned, where having a tilted plane with an incidence angle of >45$^{\circ}$ has been demonstrated to increase the track acceptance \cite{Granja2018b}. Therefore, measurements were also performed with the detector at 60$^{\circ}$ in the perpendicular plane. 

\begin{figure}[!htb]
    \centering  
    \includegraphics*[width=0.52\columnwidth]{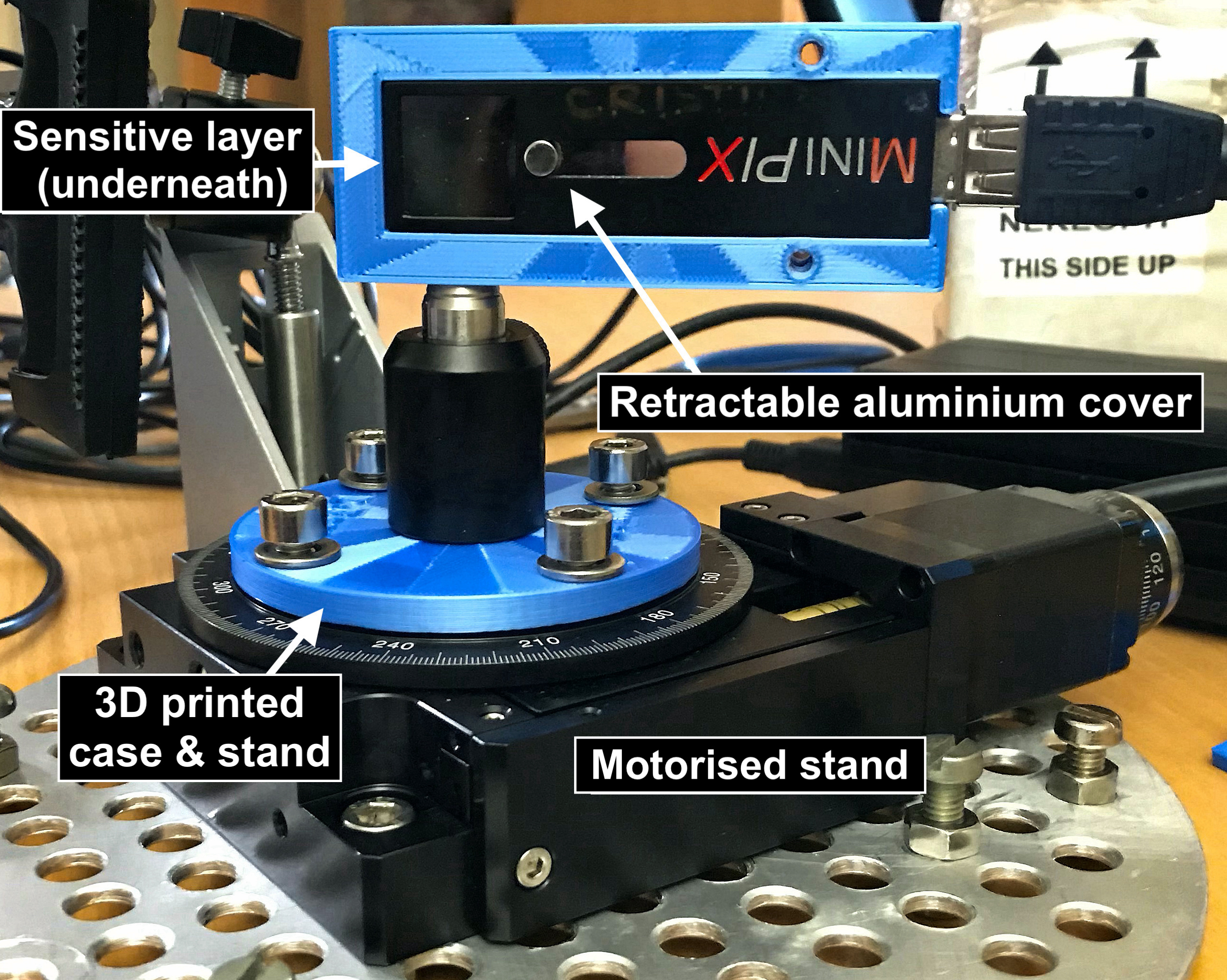}
    \caption{The MiniPIX detector secured to a motorised, rotating base. A retractable 1~mm aluminium sheath covers the chip sensitive area and the centre of the sensor is positioned along the origin of the rotational axis.}
    \label{fig4}
\end{figure}

The detector was mounted in a custom 3D printed case and attached to a remotely controlled, motorised rotating stand which was positioned perpendicular to the propagation direction of the beam (figure \ref{fig4}). This was bolted to a metal plate and clamped to the CCC treatment chair for all experimental runs. Different acquisition times ($\sim$1 min) were used to obtain sufficient statistics with the detector angled at 45$^{\circ}$ and 60$^{\circ}$ in the perpendicular plane. The detector was positioned at the same distance (approximately 18~cm) from the nozzle for all the runs. 

Wide ranging measurements of both heavy and light charged particles (protons, electrons etc.) are enabled through position and directional tracking for each pixel, combined with the timing and energy detection capabilities. Given this, it is possible to determine particle specific properties including the type, energy loss, LET, track mapping and also beam related properties such as profile, flux and dose rates. The vendor supplied Advacam PIXet Pro readout software \cite{Turecek2015} was used to record data in real-time. How the software incorporates and analyses this data is not detailed in this paper but is described in \cite{Granja2018b}. For our measurements, the energy loss, elevation angle and 3D path length were considered and used to directly resolve the LET spectra. 

\subsection{Beam Measurements}
At CCC, typical treatment beam settings achieve flux rates up to 10\textsuperscript{10} protons/s at the nozzle, corresponding to a beam current of the order of nA. The cyclotron generates comparatively high currents upstream as there are large losses in beam transmission at both scattering foils given the passive delivery system. Due to the count rate capabilities of the Timepix chip, the beam current needed to be heavily reduced to ranges suited for the MiniPIX; optimally within 10\textsuperscript{3}--10\textsuperscript{4} protons/s/cm\textsuperscript{2} to prevent damage to the sensor, and event pile-up ($<$10\textsuperscript{6} protons/s/cm\textsuperscript{2}) \cite{Granja2018,Nabha2022}.

\begin{figure}[!hb]
    \centering  
    \includegraphics*[width=0.7\columnwidth]{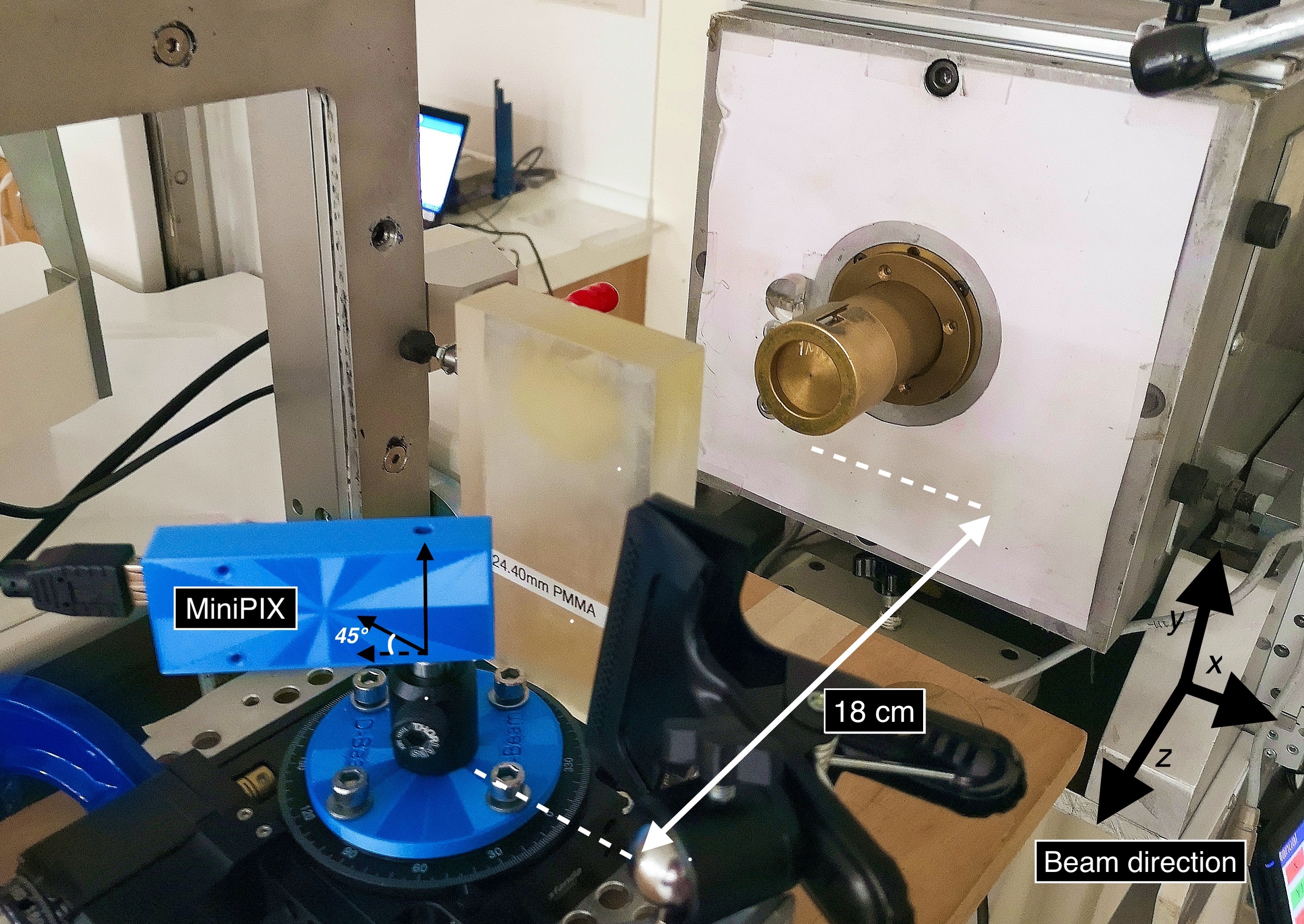}
    \caption{Experimental setup of the MiniPIX detector angled at 45$^{\circ}$ to the beam, positioned 18 cm from the collimator. Different thickness blocks of PMMA (Lucite) were placed upstream to change the WET, shifting the depth of the BP (24.40~mm block is pictured).}
    \label{fig5}
\end{figure}

The detector was firstly aligned perpendicular to the propagation direction of the beam. Several factors also needed to be considered to account for operational beam uncertainties discussed in \cite{Yap2019,Yap2021b}. Preliminary simulations in TOPAS were performed to investigate different approaches and materials to limit particle transmission, and for initial calibration. For the first runs, a 1~mm nozzle brass collimator and a lead sheet with a 200~$\upmu$m diameter pinhole was placed in front of the detector such that appropriate adjustments could be made with the accelerator to reduce the flux by several orders of magnitude, and to stabilise the beam. Following this, the nozzle collimator was retained and the detector was rotated at an angle of 45$^{\circ}$ to the beam to acquire measurements. Data was first taken over different acquisition times to select the appropriate detector settings before the different PMMA (Lucite, brand name) blocks were inserted directly upstream of the sensor to modify the depth the protons would traverse through (figure~\ref{fig5}).

Prior to recording measurements, initial digital tests were performed with the detector, showing zero dead pixels. A recommended bias of +30~V was applied with a frame acquisition time of 10~ms. A visual check is sufficient to ensure that the chosen frame rate results in frames where almost all tracks are separated from each other. As the frame rate is decreased for a fixed flux, it becomes less feasible to distinguish single particle tracks as they overlap more and more. The detector threshold was set to the minimum level, just above the noise for maximum sensitivity and set to measure ToT (Time-over-Threshold). There is a non-linear calibration from ToT to energy which is accounted for in the acquisition software. Data was taken for similar total acquisition times for each run, mainly to acquire adequate statistics. Cluster files reported global frame times every $\sim$30~ms: this consists of the 10~ms open shutter time with the remaining as dead time for readout (closed shutter), resulting in an effective frame rate of 33.3 FPS. The readout software displayed information for each frame or a visualisation of hits integrated across all frames. Along with real-time cluster statistics and analysis, these also indicated which depths were most significant relative to the LET, particularly leading up to the BP. Due to time restrictions following the 45$^{\circ}$ measurements, the 60$^{\circ}$ runs were limited to three cases (10~mm, 24.40~mm, and 25.87~mm PMMA blocks) chosen for comparison.

\section{Results}
\subsection{Water Equivalent Thickness}
As objects of different materials intercept the beam path, the total water equivalent thickness (WET) was calculated to provide a more suitable indicator of depth along the full proton range. The aluminium slider introduced another attenuating layer in addition to the various PMMA blocks and was implemented in TOPAS to accurately simulate the experimental conditions. The angular rotation of the detector also provided additional depths for measurements. All corresponding WET values are listed in table \ref{T_minipixWET}. The depth at which particles have undergone almost complete energy loss and are fully absorbed has been designated a WET of $\infty$.

\begin{table}[!hb]
\centering
\begin{threeparttable}
\caption{Experimental conditions and calculated WET for simulations.}
\label{T_minipixWET}
\smallskip
\begin{tabular}{c|c|c|c}
\hline
Detector angle {[}$^{\circ}${]} & PMMA thickness {[}mm{]} & Al effective thickness {[}mm{]} & WET {[}mm{]} \\
\hline
    45    & 10.00   & 1.4   & 14.42 \\
    60    & 10.00 & 2.0   & 15.64 \\
    45    & 20.10  & 1.4   & 26.00 \\
    45    & 24.40  & 1.4   & 30.87 \\
    45    & 25.48 & 1.4   & 31.96 \\
    60    & 24.40 & 2.0   & 32.00 \\
    45    & 25.87 & 1.4   & $\infty${*} \\
    45    & 26.40 & 1.4   & $\infty${*} \\
    60    & 25.48 & 2.0   & $\infty${*} \\
\hline
\end{tabular}
\begin{tablenotes}
  \item[*]A WET of $\infty$ indicates that the vast majority of protons have stopped within the material and therefore do not reach the detector.
  \end{tablenotes}
\end{threeparttable}
\end{table} 

\subsection{Particle Track Clustering}
\begin{figure}[!htb]
    \centering  
    \includegraphics*[width=0.75\columnwidth]{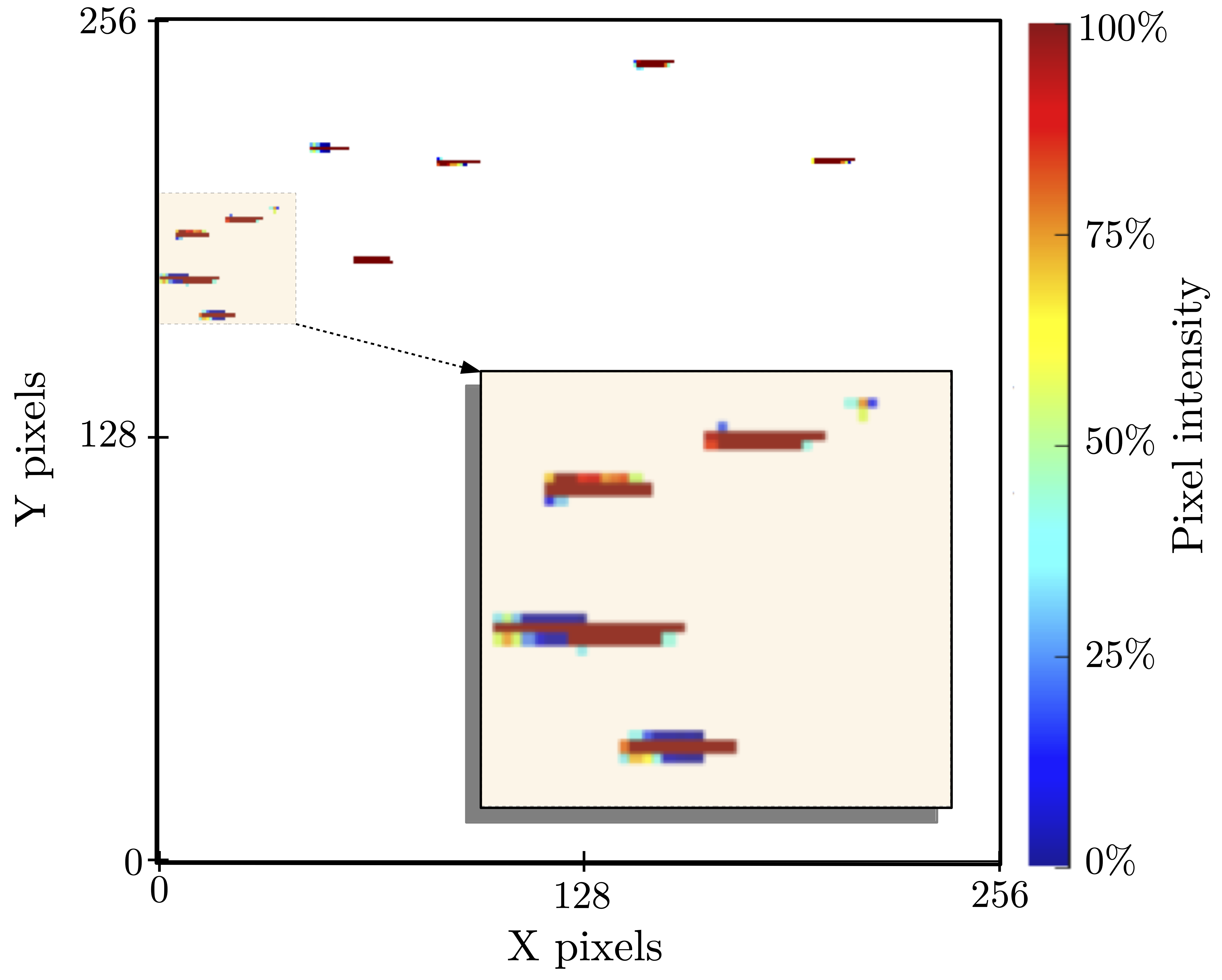}
    \caption{Visualisation of individual clusters shown using the Pixet propietary readout software. Charge deposition is shown given relative pixel intensities, acquired in single frame (10~ms) for R7 (table \ref{T_minipixRecord}).}
    \label{fig6}
\end{figure}

The detector recorded events containing the charge deposition produced by individual particles. Single particles often trigger multiple pixels resulting in a cluster, a region containing more than one triggered pixel surrounded by zero hits (figure \ref{fig6}). The readout software has inbuilt clustering algorithms to identify and classify clusters based on morphology, spectral, and other parameters \cite{Granja2018}. Different tools also allow further online or offline processing however standard settings and the energy calibrations provided by the vendor were used throughout the experiment and for analysis. 

Using the readout software, log files of data were processed to generate various metrics (energy deposited, track sizes, angle etc.) for the distributions of events for each run, stored as lists of cluster properties. The raw cluster statistics and details of each measurement are shown in table \ref{T_minipixRecord}.

\begin{table}[!ht]
\centering
\begin{threeparttable}
\caption{Measurement details and simulation labels corresponding to each run.}
\label{T_minipixRecord}
\smallskip
\begin{tabular}{c|c|c|c|c|c}
\hline
Run [\#] / & Detector & WET & Acquisition & Number of & Cluster rate \\
Simulation [\#] & angle {[}$^{\circ}${]} & [mm] & time [s] & clusters & [clusters/s] \\
\hline
R1 / S1a & 45 & 14.42 & 131.35 & 9440  & 693.9 \\
R2 / S2a & 45 & 26    & 68.07  & 7277  & 909.3 \\
R3 / S3a & 45 & 30.87 & 58.44  & 3445  & 581.1 \\
R4 / S4a & 45 & 31.96 & 58.44  & 2270  & 460.3 \\
R5 / S5a & 45 & $\infty${*}    & 58.44  & 569   & 294.3 \\
R6 / S6a & 45 & $\infty${*}    & 58.44  & 179   & 203.9 \\
R7 / S7a & 60 & 15.64 & 58.44  & 21610 & 1341  \\
R8 / S8a & 60 & 32    & 58.44  & 5325  & 693.9 \\
R9 / S9a & 60 & $\infty${*}    & 58.44  & 344   & 259.2 \\
\hline
\end{tabular}
\begin{tablenotes}
  \item[*]A WET of $\infty$ indicates that the vast majority of protons have stopped within the material and therefore do not reach the detector.
  \end{tablenotes}
\end{threeparttable}
\end{table}

Greater statistics of clusters depositing larger amounts of charge, and higher energy levels with a much bigger range of energy per cluster were observed at runs with the 24.40 mm and 25.48 mm blocks. This raw data is not presented here but is detailed in \cite{Yap_Thesis}. The increased frequency of interactions and spread of deposited energies at these depths could be seen in post-processing, corresponding with the BP. Similar trends in the cluster data for both the 45$^{\circ}$ and 60$^{\circ}$ measurements were also observed, with scaling differences between the two cases. This is due to the change in the angle of incidence by 15$^{\circ}$ which increases the sensitive area exposed, given a larger effective sensor depth. However this also increases the probability of detecting longer tracks, therefore scaling the energy deposition across the whole path length, resulting in similarly equivalent LET values. Double peaks were found in the data also indicating that the detector recorded multiple particles causing an overlapping of tracks, which may have distorted the true cluster size distribution. It is also unclear if limitations with beam stabilisation and calibration had a significant effect on device performance. However it is expected that the distributions would be more consistent with longer exposure times and more counts. Deeper cluster analysis may also better identify individual events and reduce duplication of hits, however would require technical manipulation of the software and was outside the scope of this work. 

\subsection{Linear Energy Transfer}
\begin{figure}[!ht]
\includegraphics[width=\columnwidth]{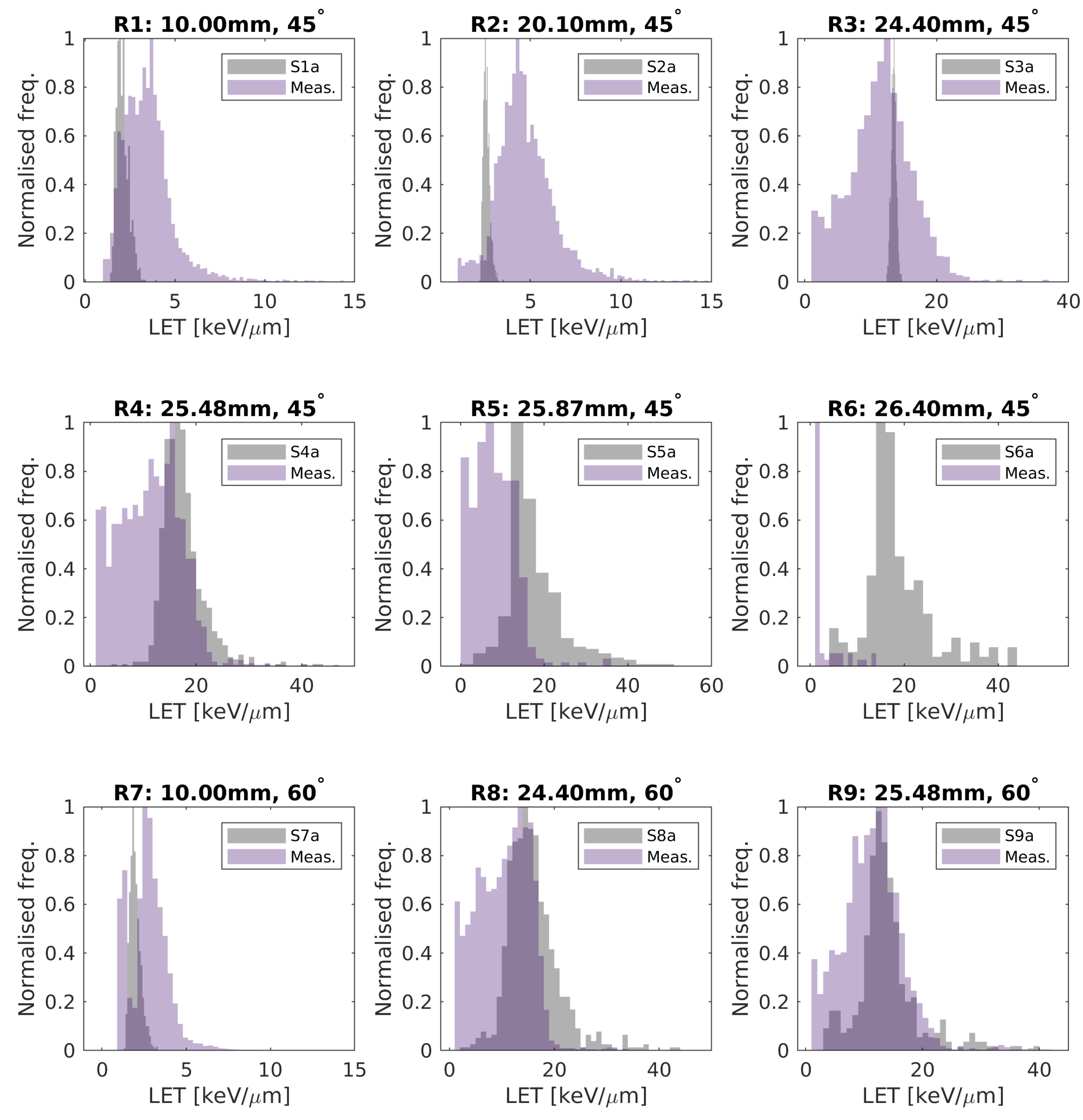}
\centering
\caption[]{LET spectra in silicon for all runs (R1--R9) derived from the MiniPIX sensor (Meas.) data compared to simulation data (S1a--S9a).} 
\label{fig7}
\end{figure}

The LET spectra for all charged particles could be determined by evaluating the cluster lists and taking the cluster volume (energy deposition, d\textit{E}) and cluster size distributions. As tracks were recorded in any orientation, the actual particle path length (track length, d\textit{l}) could be calculated given the directional angle of the incident track, sensor thickness and the measured projected length \cite{Granja2018}. For the measured energy deposition spectra, the detected energies required further processing from the cluster volumes to enable direct comparison with the simulated data. The experimental setup was simulated with each particle scored at entry to the simulated sensor volume, generating a spectrum of kinetic energies \cite{MarkThesis}. The empirical data was compared to the simulation data by interpolating the stopping power in silicon using the NIST PSTAR database \cite{NISTepSTAR} for all kinetic energies. These were then multiplied by the recorded path lengths to obtain the detector measured energy deposition data. These quantities were used to evaluate the LET values as given by equation \ref{eqn1}, resulting in a spectra of LET values (figure \ref{fig7}). 

\begin{figure}[!ht]
\includegraphics[width=\columnwidth]{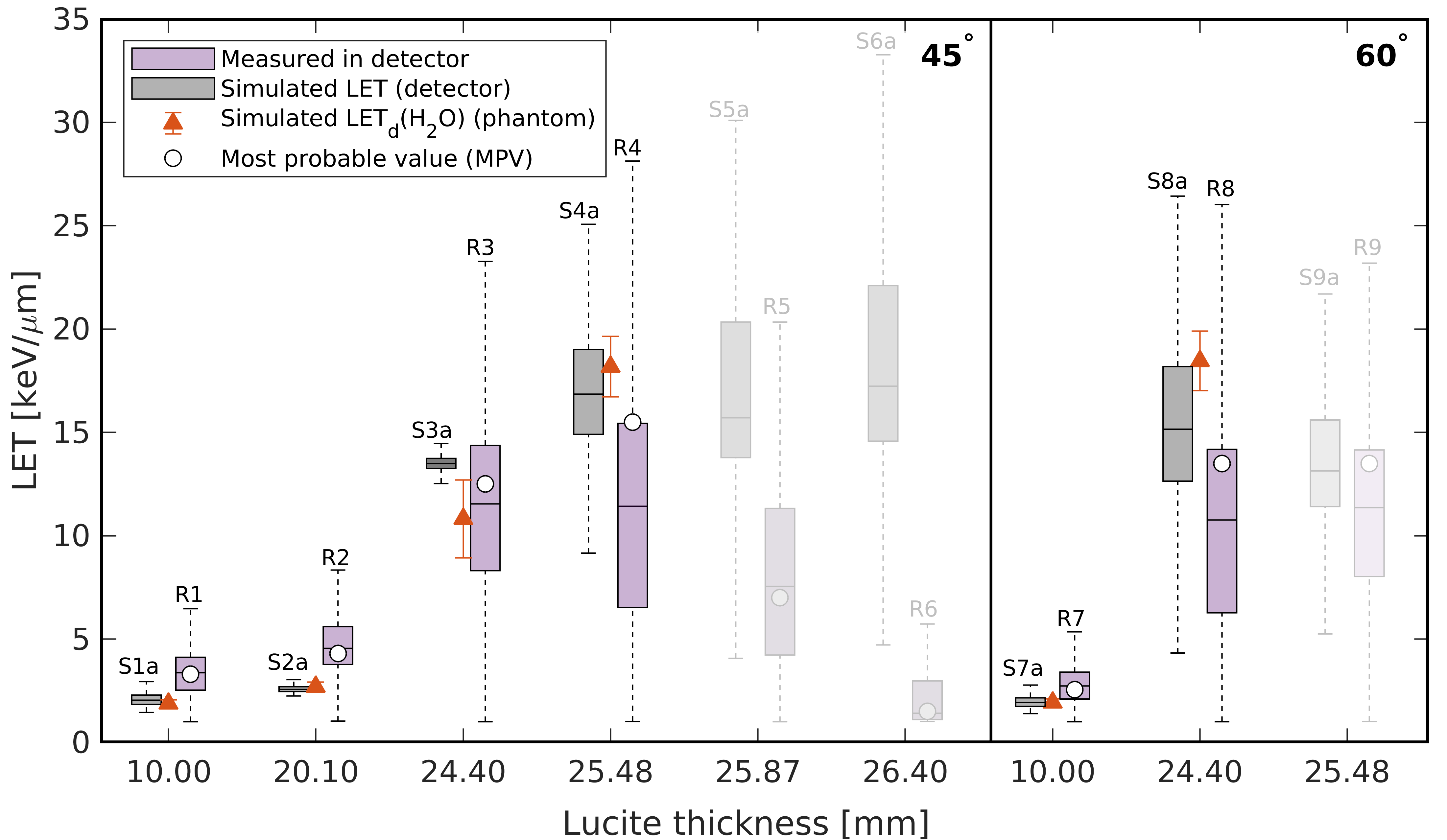}
\centering
\caption[]{Box plots of the measured LET data with the detector for all runs (R1--R9) with different thicknesses of PMMA (Lucite), at 45$^{\circ}$ and 60$^{\circ}$ and compared with simulated cases (S1a--S9a). MPVs determined by Landau fits to the spectra show the indicative measured LET values (for all particles). A standard simulated LET\textsubscript{d} water phantom case is also displayed as reference. Cases with a designated $\infty$ WET are shaded.} 
\label{fig8}
\end{figure}

Given the spectra of LET values at each depth demonstrated by both empirical and simulation methods, a Landau \cite{Meroli2011} or other stable distributions could be fitted to the data \cite{MarkThesis}. These produce a peak: the most probable value (MPV), chosen as indicative of a median or average metric to represent the LET. The full range of LET spectra values simulated and empirically measured by the detector (for all particles) is presented in figure~\ref{fig8} as box plots for all depths and angles. We show the LET distributions in silicon given the detector sensor material but for guidance, we also include a standard simulated case with a water phantom, using the preset LET\textsubscript{d} TOPAS scorer. Detailed conversion between the different materials is outside the scope of this work.

Figure \ref{fig8} shows a large range of LET values, increasing up to the BP and beyond. There is a dominance of events at the lower end of the LET spectra partly due to the presence of the 1~mm thick aluminium slide cover, inducing small peaks and events at the sensor entrance. The aluminium would have effectively functioned as a high-pass filter, as protons \cite{NISTepSTAR} less than $\sim$13~MeV would have been absorbed before reaching the sensor. However, this may have produced additional low energy, high-LET particles contributing to the spread of values. The cover was kept closed as a protective measure and remained shut for the remainder of the measurements.  

\begin{figure}[!ht]
\includegraphics[width=\columnwidth]{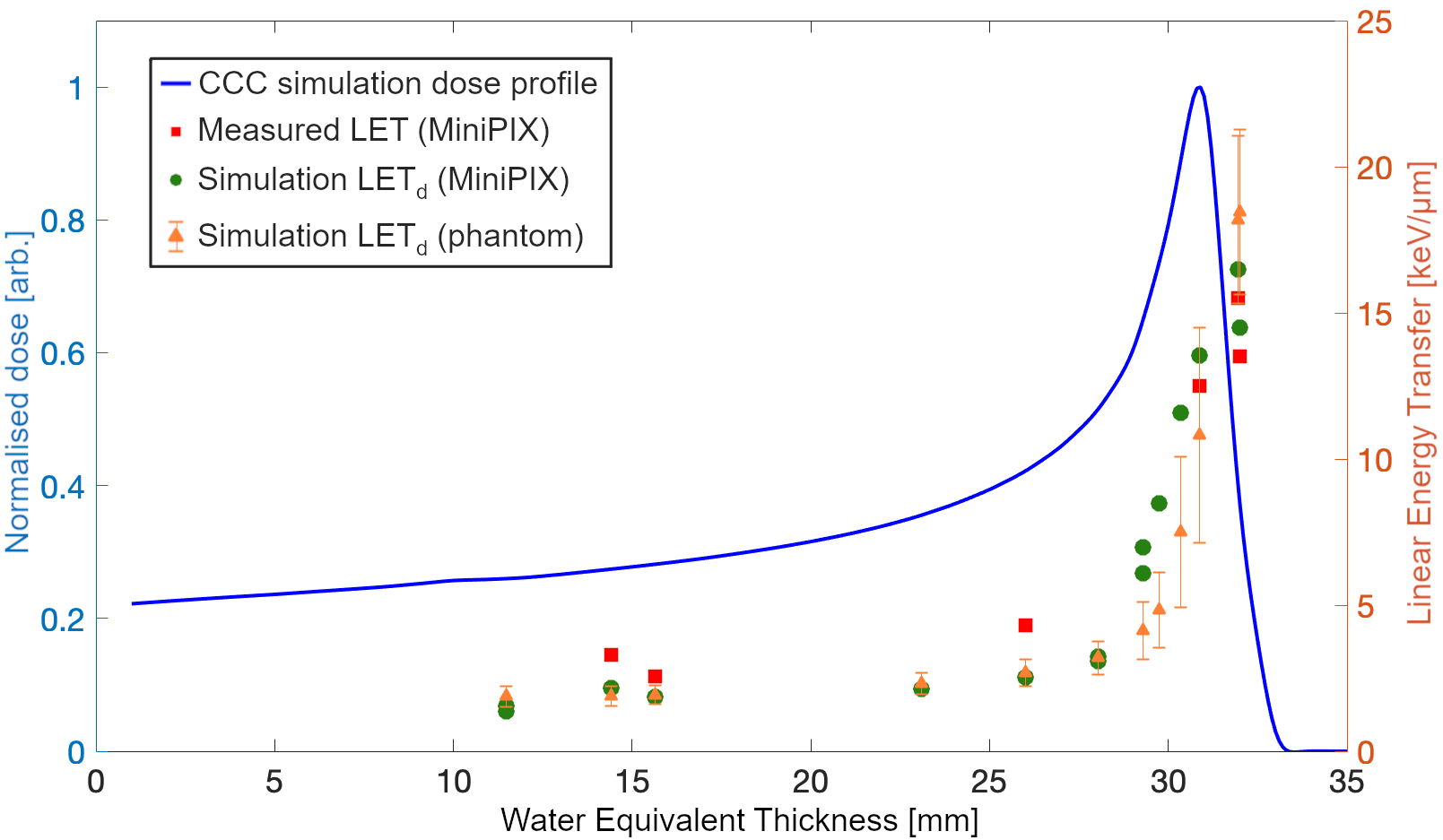}
\centering
\caption[]{Measured LET values with the MiniPIX detector compared with the dose averaged LET simulated under experimental conditions (right axis). The simulated CCC dose profile (left axis) and standard LET\textsubscript{d} cases (in water) are shown for reference.} 
\label{fig9}
\end{figure}

As the irradiations covered a wide range of depths particularly around the BP region, the LET is also shown with the simulated dose profile against WET in figure \ref{fig9}. The previously obtained MPVs in figure \ref{fig8} are shown here as the measured MiniPIX LET data. The experimental case simulated in TOPAS and standard LET\textsubscript{d} reference case is also presented, error bars are included to indicate the variance reported by TOPAS. 

Figure \ref{fig9} shows that the measurements are within range of the simulations and expected LET values. As seen for all cases, the linear energy transfer is maintained below approximately 4~keV/$\upmu$m before showing a clear upward trend from the BP build-up region to the distal fall-off. Following this, the increased density of ionisation and nuclear interactions cause the protons to slow dramatically; the transfer of energy to secondary particles results in high deposition events in short ranges, demonstrated by the steep rise at the BP. An LET of $\sim$12~keV/$\upmu$m is obtained at the BP. Past the peak, the fluence rapidly reduces as protons deposit the majority of their energy downstream, achieving maximum LET values between 13.5--18.2~keV/$\upmu$m. Larger errors are observed at the fall-off due to statistical fluctuations and the departure of primaries which generate electrons or other secondary particles: these are scattered with a wide range of energies and path lengths. After this point all particles have undergone complete energy loss resulting in no further LET.

The simulated LET\textsubscript{d} cases are shown to be in reasonable agreement despite the difference in absorption material, particularly at depths up to the build-up region. However, as the number of interaction events increases, the LET becomes more difficult to model and the deviation between the data sets is clear. The number of simulated particles and recorded MiniPIX events at the terminal end (past the BP) are likely to be statistically insufficient. At the BP, the simulated and measured MiniPIX LET values are larger than for the reference phantom case, except at 32~mm WET where they are mostly contained within the uncertainty range. The greater LET values building up to the BP may be correlated to the differences between stopping power with silicon and water but encourages additional measurements and further investigation around the WET. At the fall-off, the distribution (interquartile range and median) of measurements skew toward lower energies (figure \ref{fig8}), hence result in smaller empirical LET values when compared to simulations.

In the plateau region before the BP, the empirical data has tended to overestimate the LET. As mentioned previously this is largely attributed to the presence of the aluminium slider and complexities with the analysis of the clusters. For example, overlapping events recorded by the detector produced multiple peaks in the cluster distributions for the runs using PMMA blockers thicker than 24.40~mm. This results in an increase in the range of track lengths and also the variance in the deposited energy per cluster, therefore a greater spread in the energy deposition. Misclassification of tracks or particle type could also result in electron contributions  being included in proton stopping power calculations and therefore measured energy deposition data. It is also unclear if limitations with preparation and beam calibration had a significant effect on device performance. It is expected that the distributions would be more consistent with increased fluence.

\begin{figure}[!ht]
	\includegraphics[width=0.9\columnwidth]{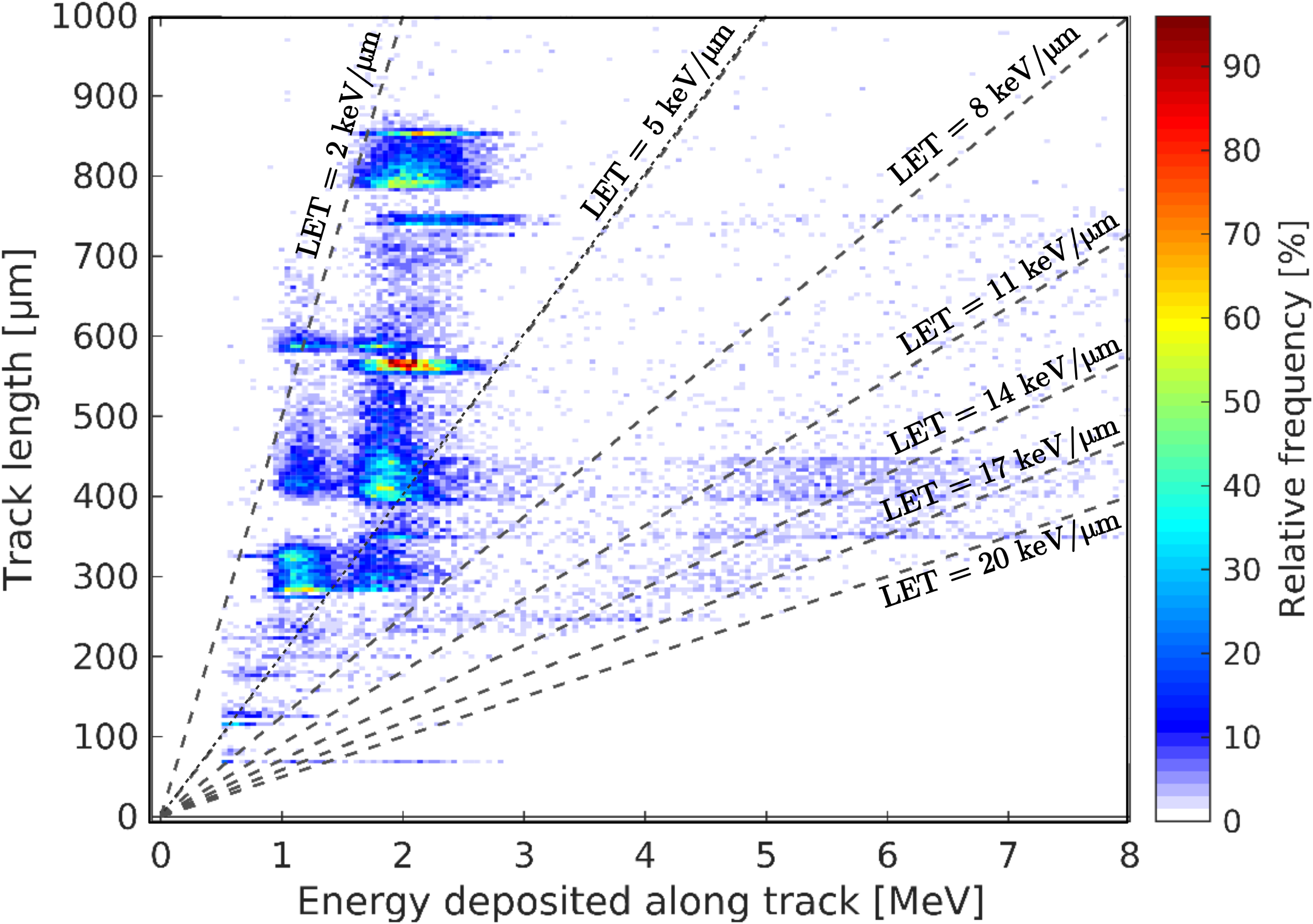}
        \centering
        \caption[]{Distribution of deposited energy and track lengths from all the measured MiniPIX data (R1–R9 conditions as listed in table \ref{T_minipixRecord}). Level curves of constant LET values are shown for reference.}
        \label{fig10}
\end{figure}

Figure \ref{fig10} provides an alternative visualisation of the MiniPIX data, a track-energy spectral plot (TESP). The TESP captures information about the energy deposition, event track lengths, and resulting LET spectra in a single plot. This shows the distribution of deposited energy and lengths of particle tracks that contribute to the calculated LET, across the different LET levels. To our knowledge, this type of data has not yet been presented in the existing literature. 

In the TESP the majority of tracks contributing to LET are grouped at the lower end of values (i.e. $<$8~keV/$\upmu$m), where tracks were largely measured from 400--600~$\upmu$m and at pockets of 300~$\upmu$m and 800~$\upmu$m. Contributions here by electron contamination are expected but the extent is difficult to evaluate. It may be possible to reduce this contamination and the incidence of non-proton events by implementing a cut-off threshold to remove small tracks \cite{Granja2021}. These contributions could be seen in the multi-peaked distributions of the cluster data where the presence of partial or complete peaks at lower energies were not completely filtered out, translating to a deviation of the LET spectra from a Landau distribution. This can be observed for many cases shown in figure \ref{fig7}. There are several differences to the simulated results as these additional electron contributions were not included. The higher energy depositions are mainly made up of tracks at 400~$\upmu$m, also contributing to some of the higher LET events. Contributions to the largest LET values, particularly at the furthest end of the distal fall-off are correlated to a group of short track events where particles undergo most of their energy loss within single pixels (55~$\upmu$m). 
 
To simulate the sensor in TOPAS, small bin sizes were selected for the detector scoring geometry however there was also the possibility of artificially overestimating LET contributions due to the voxel size being smaller than the mean path length between events. Measurements of the LET are confined to the scoring bin, instead of the track being traced beyond these limits. The scoring geometry defined for the detector simulations consisted of voxels with dimensions $0.5 \times 0.5 \times 0.3$~mm$^{3}$. The largest possible track length within a voxel was therefore 768~$\upmu$m, corresponding to the longest corner-to-corner distance. Furthermore, the angle of the detector in the simulations could have allowed a portion of particle tracks to traverse one voxel but then leak into another. This second voxel would then record a higher LET track that has already lost energy when passing through the first voxel which can also be a source of overestimation. More detailed filtration and post-processing techniques to remove artifacts and other contamination may help to better identify tracks and particle types, precise energy deposition contributions and therefore the determination of LET in future.

\section{Conclusion}
Measurements were performed with a MiniPIX-Timepix detector at the Clatterbridge Cancer Centre 60~MeV proton therapy beamline to determine the LET along the proton range. The detector was positioned at two different angles to change the proton incidence angle, and PMMA blocks of various thicknesses were used to change the WET and corresponding BP depth. We developed a precise, end-to-end model of the CCC beamline in the Monte Carlo code TOPAS, simulating the realistic experimental conditions of the clinical beam. Measured cluster data provided the energy deposition and track structure information necessary to evaluate the LET spectra, and determine appropriate values for comparison. The measured LET values were shown to generally agree with simulated data; variations are likely caused by experimental considerations and differences in calculation methods. Additional cluster processing and analytical methods were outside the scope of this study but were discussed, and could enhance the determination of tracks and deposition events to improve upon measured data. In practice it is difficult to determine the LET as no commercial tools are readily available which can provide detailed, direct measurements. In this paper we demonstrate the capabilities of the MiniPIX detector to determine precise quantities of the track length and deposited energy to resolve the linear energy transfer, supported by simulation modelling. This work contributes to further development of this technology for use in proton therapy and possible applications where LET quantities are important for radiobiological studies or patient treatment planning. 

\acknowledgments
The authors would like to thank Matthieu Hentz, Ohie Mayenin and Raffaella Rodogna from UCL for their contributions toward developing the CCC simulation model. Some of this work has been presented previously in PhD theses \cite{Yap_Thesis,MarkThesis} and a conference proceeding \cite{Yap2021a}. The authors have no relevant conflicts of interest to disclose. This research was funded by the European Union FP7 grant agreement 215080, Horizon 2020 research and innovation programme: Marie Sk\l odowska-Curie grant agreement no. 675265 -- Optimization of Medical Accelerators (OMA) project and the Cockcroft Institute core grant STGA00076-01. The work of Mark Brooke was supported by Cancer Research UK, grant number C2195/A25197, through a CRUK Oxford Centre DPhil Prize Studentship. The work of Carlos Granja at VSB TU Ostrava was supported by the project SGS SP 2024/016 at the VSB Ostrava financed by the Ministry of Education, Youth and Sports of the Czech Republic.

\bibliographystyle{JHEP}
\bibliography{References}






\end{document}